# First principles study of magnetism in nanographenes

*De-en Jiang[1,*], Bobby G. Sumpter[2,3], Sheng Dai[1,3]*

[1]Chemical Sciences Division and [2]Computer Science and Mathematics Division and [3]Center for Nanophase Materials Sciences, Oak Ridge National Laboratory, Oak Ridge, Tennessee 37831

Abstract: Magnetism in nanographenes (also know as polycyclic aromatic hydrocarbons, or PAHs) are studied with first principles density functional calculations. We find that an antiferromagnetic (AFM) phase appears as the PAH reaches a certain size. This AFM phase in PAHs has the same origin as the one in infinitely long zigzag-edged graphene nanoribbons, namely, from the localized electronic state at the zigzag edge. The smallest PAH still having an AFM ground state is identified. With increased length of the zigzag edge, PAHs approach an infinitely long ribbon in terms of (1) the energetic ordering and difference among the AFM, ferromagnetic (FM), and nonmagnetic (NM) phases and (2) the average local magnetic moment at the zigzag edges. These PAHs serve as ideal targets for chemical synthesis of nanographenes that possess magnetic properties. Moreover, our calculations support the interpretation that experimentally observed magnetism in activated carbon fibers originates from the zigzag edges of the nanographenes.

*To whom correspondence should be addressed. E-mail: jiangd@ornl.gov. Phone: (865) 574-5199. Fax: (865) 576-5235.

**I. Introduction**

Graphene-based materials have attracted great attention since the successful isolation of graphene, a single layer of graphite, in 2004.[1-3] This top-down approach to graphene preparation enabled a plethora of experimental studies to probe the unique properties of electrons in the two-dimensional honeycomb lattice.[4-8] Fascinating phenomena such as chiral quantum Hall effect have been observed.[9] Moreover, graphene-based electronics may provide a new generation of nanoscale devices.[9] In addition, graphene may be particularly good for composite materials[10] as well as a number of other exciting applications.[9]

Edge effects will become significant when a graphene layer is shrunk to nanometer scale for electronic devices such as transistors and logic gates.[11] Two main types of edges exist in nanographenes: armchair and zigzag. The properties of zigzag edges have been reported as earlier as in 1993 when two theoretical papers predicted localized electronic states at the zigzag edges.[12,13] Later, detailed theoretical studies predicted an antiferromagnetic (AFM) ground state for zigzag-edged graphene nanoribbons (ZGNR).[14,15] Recently, unique physical and chemical properties of ZGNRs from first principles density functional calculations have been reported.[16,17] In the meantime, the localized electronic states predicted for ZGNRs were confirmed by scanning tunneling microscopy and spectroscopy.[18,19]

Magnetism in nanographites and nanographenes can also arise from vacancies, dopants, curvatures, and proton irradiation.[20-23] However, here we focus our discussion on edge effects of π-electrons on magnetism of nanographenes (all carbon atoms in $sp^2$ state and dangling σ-bonds saturated by hydrogen). Although experimental work has been scarce, quite a few theoretical reports have been published.[14-16,24-34] ZGNRs have been shown to have an AFM ground state with one edge spin up and the other spin down. Following this AFM phase in stability are a ferromagnetic (FM) state and then a nonmagnetic (NM) state. The armchair-edged graphene nanoribbons (AGNRs) do not show such magnetic phases. Although the AFM phase has not been reported for a single nanographene ribbon, Enoki and coworkers have observed localized spins in nanographite domains of activated carbon fibers and attributed the origin to the zigzag edges.[35,36]

Before experimental verification, ZGNRs with reasonable length that can preserve magnetic and electronic properties predicted for infinitely long ribbons need to be synthesized. Chemists have made great stride in the bottom-up synthesis of nanographenes.[37] However, synthesis of rectangular nanographenes with consecutive zigzag edges remains a challenge.[38-40] Nanographenes are also known as polycyclic aromatic hydrocarbons (PAHs). Fig. 1 defines a rectangular PAH [X,Y]: X and Y represent the number of fused rings in the zigzag and armchair edges, respectively.[41] First principles studies of magnetism have been reported mainly for infinitely long ribbons, and to the best of our knowledge, detailed examination of magnetism in nanometer-sized PAHs has not been attempted from first principles. In the present work, we would like to address the following two questions with first principles density functional theory (DFT) calculations: (1) how do the magnetic properties of rectangular PAHs change with sizes and how do they compare with infinitely long ribbons; and (2) how small can a PAH be but still have a magnetic ground state? Answering these questions will help understand the magnetism of nanographenes and nanographites and also guide synthesis towards a small enough PAH that still behaves electronically like a long ribbon.

## II. COMPUTATIONAL METHODS

The Vienna Ab Initio Simulation Package (VASP)[42,43] was used to perform DFT calculations with planewave bases and periodic boundary conditions and within the generalized-gradient approximation (GGA) for electron exchange and correlation.[44] Projector-augmented wave (PAW) method[45,46] was used within the frozen core approximation to describe the electron-core interaction. A kinetic energy cutoff (450 eV) was used. Supercell models were employed; i.e., a PAH molecule was put in a large box. Because PAH molecules in the present work are flat and rectangular, the molecules were placed in the *xy*-plane. The *x* and *y* dimensions of the boxes range from 15 Å by 15 Å for PAH [3,3] to 30 Å by 20 Å for PAH [9,5]. The *z* dimension for all the boxes was fixed at 10 Å. Only the Γ-point was used to sample the Brillouin zone. All atoms in the unit cell were allowed to relax and the force tolerance was set at 0.025 eV/Å. Full relaxation of magnetization was performed for spin-polarized calculations. All PAH molecules studied in the present work have even number of electrons, so magnetism due to odd number of electrons is not a concern here. A Voronoi scheme was employed to partition charges to atoms.[47]

In addition, we used all-electron broken symmetry DFT with an atom-centered basis (6-311G**) and hybrid exchange-correlation functionals (B3LYP)[48] and PBE0[49] to verify the energetics and existence of the magnetic phases. The main reason for these calculations was to check if there were any strong electron correlation effects (by adding exact exchange to help reduce the self interaction error) and also the effects of using a truncated planewave basis (described above) for a finite-sized molecular system. As a more rigorous test we employed many-body-wavefunction-based calculations (CASSCF) with a modest active space of 10 orbitals and 10 electrons (5 HOMOs and 5 LUMOs) using NWChem.[50] These calculations verified for PAH[4,3] the existence of a singlet antiferromagnetic phase as the ground state.

**III. Results and discussion**

We start with examining the energetics of three magnetic phases (AFM, FM, and NM)[51] for different sizes of PAHs. Table I displays the data for PAHs [3,3] to [9,5]. We find that PAH [3,3] (chemical name, bisanthene) does not have a magnetic phase; initial guesses for AFM and FM phases both converged to the NM phase. This is also the case for PAH [2,3] (perylene) and PAH [4,1] (tetracene) (results not shown in Table I). As PAH [X,Y] increases either side of the rectangle greater than 3, a stable AFM phase appears for all PAHs considered here. So does an FM phase. The AFM phase is found to be the most stable for PAHs larger than [3,3]. The FM phase is energetically in between the AFM and NM phases for most larger PAHs, except for [4,3] and [3,5] whose NM phase is more stable than the FM phase. The closeness in energy between AFM and NM phases for PAHs [4,3] and [3,5] indicates that they are in a transition between a nonmagnetic PAH [3,3] and larger PAHs that have more stable AFM phases and metastable FM phases. A good indication of how PAH [X,Y] approaches infinitely long zigzag ribbons in terms of stability is to compare the energetic difference between the AFM and NM phases (normalized to the dimension of the zigzag edge, X) of a PAH with that of an infinitely long ribbon of the same width (Y). Fig. 2 shows that the normalized energetic difference enlarges greatly from PAH [3,5] to PAH [5,5]. The change is relatively small from PAH [5,5] to PAH [9,5] and the magnitude is only slightly smaller than that of the infinite ribbon.

Although PAH [4,3] has an energetically unfavorable FM phase and a slightly more stable AFM phase, the distribution of its spin densities is prototypical for rectangular PAHs considered in the present work. Fig. 3 plots the isosurfaces of spin density magnetization ($\rho_\uparrow - \rho_\downarrow$) for the AFM and FM phases of PAH [4,3]. One can see that magnetization mainly localizes at the periphery of the PAH structure and concentrates at the middle of the two zigzag edges. In the AFM phase, the two zigzag edges have opposite spins. We note that these results agree with the all-electron atom-centered DFT and many body calculations. Another feature is that some minor magnetization also appears at the armchair edges; but as we will show later, this magnetization does not extend to the middle of the armchair edges when dimension of the armchair edge increases.

Magnetism in ZGNRs has been studied with the Hubbard model with unrestricted Hartree-Fock approximation[14] and first principles density functional theory previously.[29,34,52] Compared with an infinitely long ZGNR, the large PAHs considered here have the same stability trend, i.e., AFM < FM < NM.[16,34,52] This same trend confirms that magnetism in PAHs as patches of graphenes (or nanographenes) also arises from the zigzag edges. From Table II one can see that the total spin up magnetic moment of the AFM phase increases almost linearly with the dimension of the zigzag edge for PAH [X,5]. This increase also correlates with the stability of the AFM phase relative to the NM phase (Table I). Figs. 4a and 4b plot the spin density magnetization for the AFM phase of PAHs [7,5] and [9,5], respectively. Like in PAH [4,3], spin up electrons are localized at one zigzag edge and spin down at the other and the magnetic moments are smaller in the ends than in the middle. One can also see that the middle carbon atoms of the armchair edges have no magnetic moments, indicating that only certain armchair atoms next to the zigzag edges have significant magnetization. Average local magnetic moment for the carbon atoms at the zigzag edges is determined to be 0.172, 0.182, and 0.197 $\mu_B$ for PAHs [5,5], [7,5], and [9,5], respectively, approaching that of an infinitely long zigzag ribbon of the same width (0.22 $\mu_B$).

Unlike the AFM phase where the total spin up moment increases almost linearly with the dimension of the zigzag edge, the FM phase of PAHs displays a jump in both the total spin up moment and the total magnetization from PAH [7,5] to PAH [9,5]. The total spin up moment jumps from 2.56 to 4.96, while the total magnetization jumps from 2 to 4. For PAHs [X<9, Y], their FM phases are all triplet ($M_{total-FM}$=2, S=1).[53] Figs. 4c and 4d plot the spin density magnetization for the FM phases of PAH [7,5] and [9,5], respectively. The difference between the two is clearly visible. For PAH [9,5], the FM and AFM phases have similar distribution of magnetization except that spin is flipped in one zigzag edge. However, the FM phase of PAH [7,5] has reduced magnetization at its zigzag edges, in contrast with its AFM counterpart (Fig. 4a). We also examined the FM phase of PAH [7,5] in different magnetizations ($M_{total-FM}$=3, S=3/2, and $M_{total-FM}$=4, S=2) and found that their energies are ~100 meV less stable than the triplet.

Experimentally, Enoki and coworkers have demonstrated magnetism in activated carbon fibers.[35,36] They measured magnetization curves (susceptibility vs. applied magnetic field) down to 2 K and found that the nanographites (3~4 stacks of nanographene layers with an in-plane dimension of 2~3 nm) show the Curie-Weiss type of magnetization behavior. They deduced that each nanographite domain has 1~2 localized spins and attributed these localized spins to the localized states at the zigzag edges. Our finding that large PAHs (in-plane dimension ranging from 1 to 3 nm) have an antiferromagnetic ground state with localized magnetization at zigzag edges strongly supports their interpretation. Moreover, the number of localized spins they observed is also in line with what we found for PAHs. It should be noted that the interlayer and inter-nanographite interactions of electrons they have proposed to explain the observed magnetization curves are beyond the scope of the present work and further theoretical studies are warranted.

Recent theoretical studies have predicted many fascinating properties associated with ZGNRs, such as half-metallicity.[16] However, these predictions are mainly based on infinitely long ribbons. Before experimental verification of these predictions, ZGNRs that have reasonable length and similar magnetic and electronic properties need to be synthesized. Our results indicate that PAHs such as [5,5] preserve the properties of infinitely long ribbons of the same width. Therefore, these molecules should be good candidates for testing predictions such as half-metallicity. Although synthetic chemistry of PAHs have made tremendous progress, synthesis of PAHs such as [4,3] remains a challenge due to their instability and low solubility in common solvents.[38-40] Encouragingly, Roberson et al.[54] have observed PAHs [5,3] and [5,5] from the disproportion reaction of sublimed pentacene, using mass spectrometry. Moreover, synthesis of PAHs such as [4,3] is currently being pursued by some groups.[40]

**IV. Summary and conclusions**

We have studied magnetism in nanographenes (also know as polycyclic aromatic hydrocarbons, or PAHs) with first principles density functional calculations within the generalized-gradient approximation for electron exchange and correlation. We have shown that an antiferromagnetic phase appears at the zigzag edges as the PAH reaches a size of [4,3]. This AFM phase in PAHs has the same origin as the one in infinitely long zigzag-edged graphene nanoribbons, namely, from the localized electronic state at the zigzag edge. We found the smallest PAH that can still sustain the AFM phase to be PAH [4,3], though its energetic difference between the AFM phase and the nonmagnetic phase is small. For larger PAHs with increased dimension of the zigzag edge, the energetic ordering among the AFM, ferromagnetic (FM), and nonmagnetic (NM) phases, the normalized energetics, and the average local magnetic moment at the zigzag edges are all found to approach those of an infinitely long zigzag-edged ribbon. These PAHs are good candidate molecules for experimentally verifying predicted properties for infinitely long ribbons. Moreover, the magnetization predicted from the present first principles calculations for these PAHs (as building blocks for nanographites) supports the interpretation that observed magnetism in activated carbon fibers (containing nanographites) originates from the zigzag edges.

**Acknowledgement**

This work was supported by Office of Basic Energy Sciences, U.S. Department of Energy under Contract No. DE-AC05-00OR22725 with UT-Battelle, LLC, and used resources of the National Center for Computational Sciences at Oak Ridge National Laboratory.

FIG. 1. Defining a rectangular polycyclic aromatic hydrocarbon, PAH [X,Y]. Here $X=4$ and $Y=3$ indicate the lengths of zigzag and armchair edges, respectively. Carbon atoms are in black and H atoms in white.

FIG. 2. Change of normalized energetic difference between the AFM and NM phases of PAH [X,5] with the dimension of the zigzag edge, X. Corresponding value for an infinitely long ribbon is also plotted.

FIG. 3. Isosurfaces of spin density magnetization ($\rho_\uparrow - \rho_\downarrow$) for the AFM (a) and FM (b) phases of PAH [4,3]. Dark and light isosurfaces are 0.075 and -0.075 e/Å$^3$, respectively. Only C-C bonds are shown (C-H bonds not shown).

FIG. 4. Isosurfaces of spin density magnetization ($\rho_\uparrow - \rho_\downarrow$) for: (a) PAH [7,5] AFM phase; (b) PAH [9,5] AFM phase phases; (c) PAH [7,5] FM phase; (d) PAH [9,5] FM phase. Dark and light isosurfaces are 0.075 and -0.075 e/Å$^3$, respectively. Only C-C bonds are shown (C-H bonds not shown).

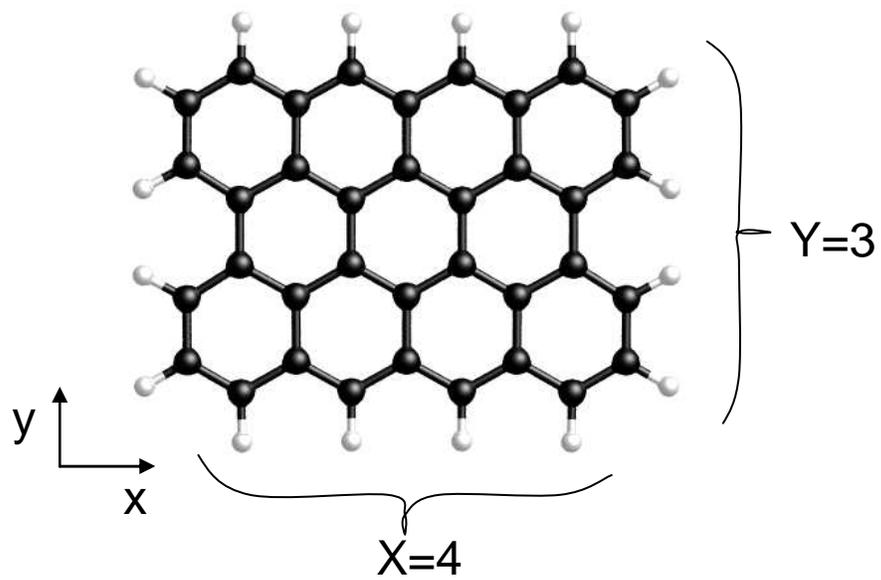

Figure 1, Jiang et al., Journal of Chemical Physics

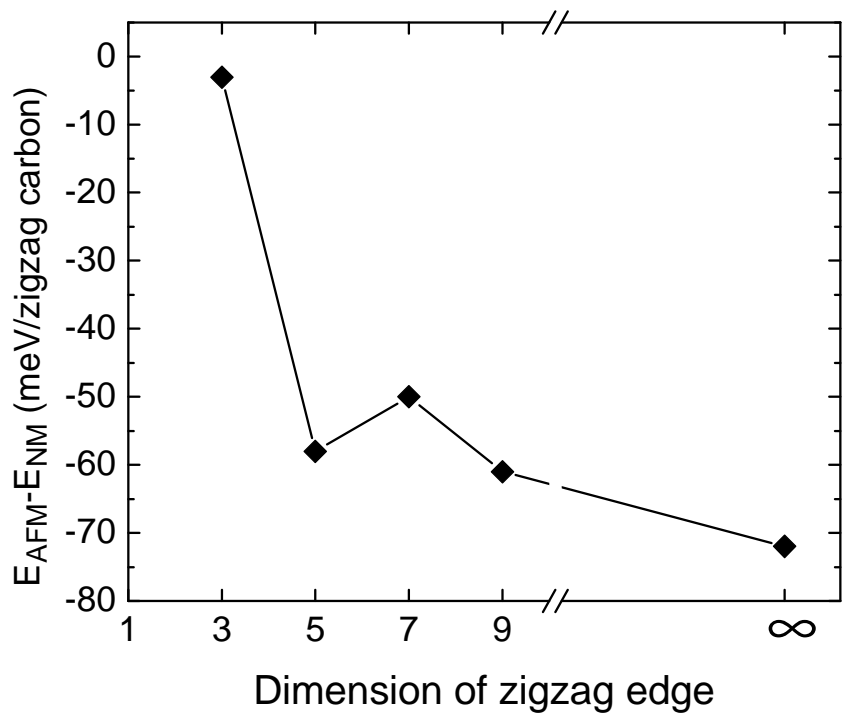

Figure 2, Jiang et al., Journal of Chemical Physics

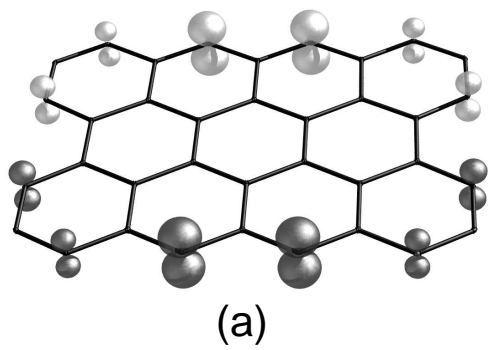 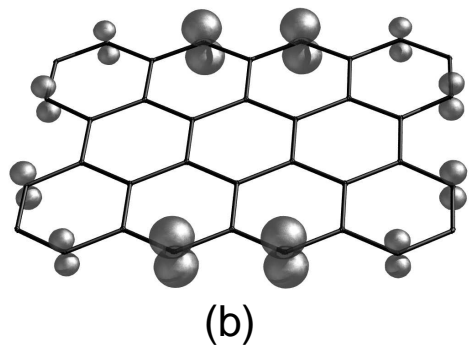

(a) (b)

Figure 3, Jiang et al., Journal of Chemical Physics
.

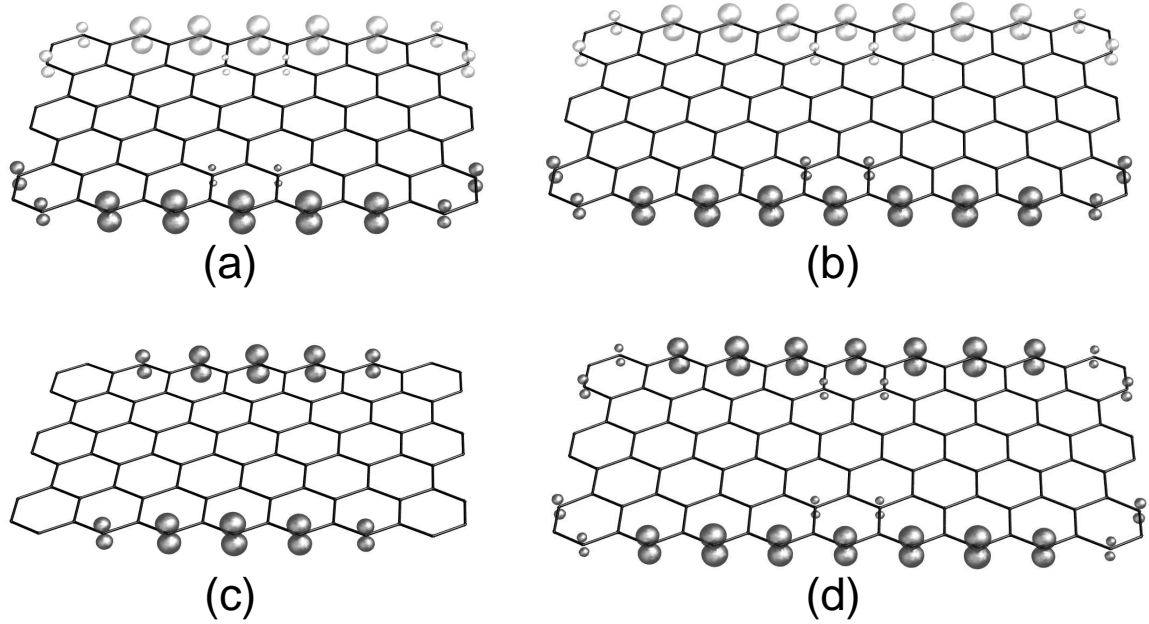

Figure 4, Jiang et al., Journal of Chemical Physics

TABLE I. Relative energies of antiferromagnetic (AFM), ferromagnetic (FM), and nonmagnetic (NM) phases of PAHs.

| PAH | [3,3] | [3,5] | [3,7] | [3,9] | [4,3] | [5,3] | [5,5] | [7,5] | [9,5] |
|---|---|---|---|---|---|---|---|---|---|
| $E_{FM} - E_{NM}$ (meV) | 0 | 102 | -88 | -173 | 43 | -202 | -277 | -260 | -502 |
| $E_{AFM} - E_{NM}$ (meV) | 0 | -18 | -113 | -180 | -52 | -230 | -290 | -349 | -548 |

TABLE II. Total magnetic moments and total spin up moments for the FM phase ($M_{total-FM}$ and $M_{up-FM}$), total spin up moments for the AFM ($M_{up-AFM}$) phase, and largest local magnetic moment on the edge carbons for the FM and AFM phases ($M_{edge-FM}$ and $M_{edge-AFM}$).

| PAH | [3,3] | [3,5] | [3,7] | [3,9] | [4,3] | [5,3] | [5,5] | [7,5] | [9,5] |
|---|---|---|---|---|---|---|---|---|---|
| $M_{total-FM}$ ($\mu_B$) | 0 | 2 | 2 | 2 | 2 | 2 | 2 | 2 | 4 |
| $M_{up-FM}$ ($\mu_B$) | 0 | 2.370 | 2.481 | 2.57 | 2.281 | 2.336 | 2.521 | 2.560 | 4.960 |
| $M_{edge-FM}$ ($\mu_B$) | 0 | 0.244 | 0.242 | 0.239 | 0.252 | 0.267 | 0.268 | 0.201 | 0.252 |
| $M_{up-AFM}$ ($\mu_B$) | 0 | 1.147 | 1.575 | 1.64 | 1.348 | 1.657 | 1.763 | 2.74 | 3.562 |
| $M_{edge-AFM}$ ($\mu_B$) | 0 | 0.164 | 0.228 | 0.238 | 0.211 | 0.266 | 0.268 | 0.236 | 0.253 |

References and notes

[1] K. S. Novoselov, A. K. Geim, S. V. Morozov, D. Jiang, Y. Zhang, S. V. Dubonos, I. V. Grigorieva, and A. A. Firsov, Science **306**, 666 (2004).

[2] K. S. Novoselov, D. Jiang, F. Schedin, T. J. Booth, V. V. Khotkevich, S. V. Morozov, and A. K. Geim, Proc. Natl. Acad. Sci. U. S. A. **102**, 10451 (2005).

[3] J. C. Meyer, A. K. Geim, M. I. Katsnelson, K. S. Novoselov, T. J. Booth, and S. Roth, Nature **446**, 60 (2007).

[4] K. S. Novoselov, et al., Science **315**, 1379 (2007).

[5] K. S. Novoselov, E. McCann, S. V. Morozov, V. I. Fal'ko, M. I. Katsnelson, U. Zeitler, D. Jiang, F. Schedin, and A. K. Geim, Nat. Phys. **2**, 177 (2006).

[6] S. V. Morozov, K. S. Novoselov, M. I. Katsnelson, F. Schedin, L. A. Ponomarenko, D. Jiang, and A. K. Geim, Phys. Rev. Lett. **97**, 016801 (2006).

[7] M. I. Katsnelson, K. S. Novoselov, and A. K. Geim, Nat. Phys. **2**, 620 (2006).

[8] K. S. Novoselov, A. K. Geim, S. V. Morozov, D. Jiang, M. I. Katsnelson, I. V. Grigorieva, S. V. Dubonos, and A. A. Firsov, Nature **438**, 197 (2005).

[9] A. K. Geim and K. S. Novoselov, Nat. Mater. **6**, 183 (2007).

[10] S. Stankovich, D. A. Dikin, G. H. B. Dommett, K. M. Kohlhaas, E. J. Zimney, E. A. Stach, R. D. Piner, S. T. Nguyen, and R. S. Ruoff, Nature **442**, 282 (2006).

[11] V. Barone, O. Hod, and G. E. Scuseria, Nano Lett. **6**, 2748 (2006).

[12] K. Kobayashi, Phys. Rev. B **48**, 1757 (1993).

[13] D. J. Klein, Chem. Phys. Lett. **217**, 261 (1994).

[14] M. Fujita, K. Wakabayashi, K. Nakada, and K. Kusakabe, J. Phys. Soc. Jpn. **65**, 1920 (1996).

K. Nakada, M. Fujita, G. Dresselhaus, and M. S. Dresselhaus, Phys. Rev. B **54**, 17954 (1996).

[16] Y.-W. Son, M. L. Cohen, and S. G. Louie, Nature **444**, 347 (2006).

[17] D. E. Jiang, B. G. Sumpter, and S. Dai, J. Chem. Phys. **126**, 134701 (2007).

[18] Y. Kobayashi, K. Fukui, T. Enoki, K. Kusakabe, and Y. Kaburagi, Phys. Rev. B **71**, 193406 (2005).

[19] Y. Niimi, T. Matsui, H. Kambara, K. Tagami, M. Tsukada, and H. Fukuyama, Phys. Rev. B **73**, 85421 (2006).

[20] P. Esquinazi, D. Spemann, R. Hohne, A. Setzer, K. H. Han, and T. Butz, Phys. Rev. Lett. **91**, 227201 (2003).

[21] N. M. R. Peres, F. Guinea, and A. H. Castro Neto, Phys. Rev. B **72**, 174406 (2005).

[22] O. V. Yazyev and L. Helm, Phys. Rev. B **75**, 125408 (2007).

[23] N. Park, M. Yoon, S. Berber, J. Ihm, E. Osawa, and D. Tomanek, Phys. Rev. Lett. **91**, 237204 (2003).

[24] K. Nakada, M. Igami, and M. Fujita, J. Phys. Soc. Jpn. **67**, 2388 (1998).

[25] K. Wakabayashi, M. Fujita, H. Ajiki, and M. Sigrist, Phys. Rev. B **59**, 8271 (1999).

[26] T. Kawai, Y. Miyamoto, O. Sugino, and Y. Koga, Phys. Rev. B **62**, R16349 (2000).

[27] K. Wakabayashi and T. Aoki, Int. J. Mod. Phys. B **16**, 4897 (2002).

[28] T. Hikihara, X. Hu, H. H. Lin, and C. Y. Mou, Phys. Rev. B **68**, 035432 (2003).

[29] K. Kusakabe and M. Maruyama, Phys. Rev. B **67**, 092406 (2003).

[30] A. Yamashiro, Y. Shimoi, K. Harigaya, and K. Wakabayashi, Phys. Rev. B **68**, 193410 (2003).

[31] M. Maruyama, K. Kusakabe, S. Tsuneyuki, K. Akagi, Y. Yoshimoto, and J. Yamauchi, J. Phys. Chem. Solids **65**, 119 (2004).

[32] M. Ezawa, Phys. Rev. B **73**, 045432 (2006).

[33] Y.-W. Son, M. L. Cohen, and S. G. Louie, Phys. Rev. Lett. **97**, 216803 (2006).

[34] L. Pisani, J. A. Chan, B. Montanari, and N. M. Harrison, Phys. Rev. B **75**, 064418 (2007).

[35] Y. Shibayama, H. Sato, T. Enoki, and M. Endo, Phys. Rev. Lett. **84**, 1744 (2000).

[36] T. Enoki and Y. Kobayashi, J. Mater. Chem. **15**, 3999 (2005).

[37] J. S. Wu, W. Pisula, and K. Mullen, Chem. Rev. **107**, 718 (2007).

[38] Z. H. Wang, E. Tomovic, M. Kastler, R. Pretsch, F. Negri, V. Enkelmann, and K. Mullen, J. Am. Chem. Soc. **126**, 7794 (2004).

[39] M. Kastler, J. Schmidt, W. Pisula, D. Sebastiani, and K. Mullen, J. Am. Chem. Soc. **128**, 9526 (2006).

[40] J. S. Wu (private communication).

[41] K. May, S. Dapprich, F. Furche, B. V. Unterreiner, and R. Ahlrichs, Phys. Chem. Chem. Phys. **2**, 5084 (2000).

[42] G. Kresse and J. Furthmüller, Phys. Rev. B **54**, 11169 (1996).

[43] G. Kresse and J. Furthmüller, Comput. Mater. Sci. **6**, 15 (1996).

[44] J. P. Perdew, K. Burke, and M. Ernzerhof, Phys. Rev. Lett. **77**, 3865 (1996).

[45] P. E. Blöchl, Phys. Rev. B **50**, 17953 (1994).

[46] G. Kresse and D. Joubert, Phys. Rev. B **59**, 1758 (1999).

[47] G. Henkelman, A. Arnaldsson, and H. Jonsson, Comput. Mater. Sci. **36**, 354 (2006).

[48] A. D. Becke, J. Chem. Phys. **98**, 5648 (1993).

[49] C. Adamo and V. Barone, J. Chem. Phys. **110**, 6158 (1999).

[50] R. A. Kendall, et al., Comput. Phys. Commun. **128**, 260 (2000).

[51] Here the "ferromagnetic" phase means that the spins at the two zigzag edges are all up. We note that some of the interior carbon atoms are spin down.

[52] H. Lee, Y. W. Son, N. Park, S. W. Han, and J. J. Yu, Phys. Rev. B **72**, 174431 (2005).

[53] $S$ is the total spin.

[54] L. B. Roberson, J. Kowalik, L. M. Tolbert, C. Kloc, R. Zeis, X. L. Chi, R. Fleming, and C. Wilkins, J. Am. Chem. Soc. **127**, 3069 (2005).